# The Gaussian Many-to-One Interference Channel with Confidential Messages

Xiang He    Aylin Yener

Wireless Communications and Networking Laboratory

Electrical Engineering Department

The Pennsylvania State University, University Park, PA 16802

*xxh119@psu.edu    yener@ee.psu.edu*

April 30, 2010

**Abstract**

The many-to-one interference channel has received interest by virtue of embodying the essence of an interference network while being more tractable than the general $K$-user interference channel. In this paper, we introduce information theoretic secrecy to this model and consider the many-to-one interference channel with *confidential messages*, in which each receiver, in particular, the one subject to interference, is also one from which the interfering users' messages need to be kept secret from. We derive the achievable secrecy sum rate for this channel using nested lattice codes, as well as an upper bound on the secrecy sum rate for all possible channel gain configurations. We identify several nontrivial cases where the gap between the upper bound and the achieved secrecy sum rate is only a function of the number of the users $K$, and is uniform over all possible channel gain configurations in each case. In addition, we identify the secure degree of freedom for this channel and show it to be equivalent to its degree of freedom, i.e., the secrecy in high SNR comes for free.

This work was presented in part at International Symposium on Information Theory, ISIT 2009, July, 2009. This work is supported in part by the National Science Foundation with Grants CCR-0237727, CCF-051483, CNS-0716325, and the DARPA ITMANET Program with Grant W911NF-07-1-0028.

# I. INTRODUCTION

In a wireless environment, interference is ever-present. Traditionally, interference is viewed as a harmful physical phenomenon that should be suppressed or avoided. On the other hand, when confidentiality of transmitted information is a requirement, because interference also limits the reception capability of a potential eavesdropper, it may end up being beneficial to the intended receiver. Hence it is important to study how to manage and/or introduce interference intelligently such that its benefit for secrecy can be harvested.

In this work, we study this problem in the framework of information theoretic secrecy. Information theoretic secrecy was first proposed by Shannon in [1] and was later extended to noisy channels by [2]–[4]. In this framework, the eavesdropper is assumed to be passive and has unbounded computation power. Secrecy is measured with mutual information: A message is said to be secure from eavesdropping if the mutual information between the message and the knowledge of the eavesdropper per channel use is negligible. The focus of problems studied in this framework is to find the fundamental transmission limits of a communication channel if the message(s) must be kept secret from eavesdropper(s).

The channel model we study in this work falls into the class of (interference) channels with confidential messages. In these channels, the receiver(s) can hear more than one transmitter and hence can eavesdrop on messages transmitted by other users in addition to receiving the message intended for it. Interference channel with two users and different eavesdropper settings have been studied extensively up to date, e.g., [5]–[12]. However, the gap between the achievable rates and the best known outer bound is still unbounded except for special cases. Reference [5] identifies the secrecy capacity region of a switching interference channel with confidential messages defined therein. Reference [7] derives the inner and outer bound of a one-sided interference channel with confidential messages and shows the gap between these bounds to be within $1$ bit. Reference [8] studies two-user interference channel with an external eavesdropper and identifies the secrecy capacity region within $0.5$ bit for the very strong interference condition.

Interference channel with more than two users has also been studied. A symmetric static $K$-user interference channel is studied in [13]. Reference [14] has studied the $K$-user interference channel where all links were i.i.d. fading and sampled from a continuous distribution. Again the results in both cases are limited to achievable rates and no outer bound is known in general.

In this work, we study the $K$-user Gaussian many-to-one interference channel, where $K \geq 3$. This model without secrecy constraints has first been studied by [15] and recently by [16]. In this channel model, only the $K$th user is interfered by the other users, hence the name "many-to-one". A main motivation for studying this channel is to be able to comprehensively characterize the role of interference in this simplified $K$-user setting. Reference [15] derives the inner and outer bound for the channel model and shows that their gap is bounded by a constant which is only a function of the user number $K$. When proving the inner bound, [15] uses layered coding and, for each layer, uses the sphere shaped lattice code from [17]. Doing so aligns signals from the first $K-1$ users at the $K$th receiver, and facilitates the decoding process at this receiver. Reference [16] characterizes a set of channel gains for which the sum capacity can be achieved by treating interference as noise at all receivers and using random codes at all transmitters.

In this work, we introduce secrecy constraints to this model in order to assess the significance of interference on secrecy considering the simplest setting for which this investigation is meaningful: The $K$th user, which is interfered by the first $K-1$ users, is also an eavesdropper on the messages transmitted by the first $K-1$ users. Hence care must be exercised to ensure the secrecy of these messages. The main contribution of this work is a lower bound and an upper bound for the secrecy sum rate of this model for *all channel gain values and interference regimes*. For two cases, we show the gap between these two bounds is only a function of the number of users $K$: (i) all channel gains between the first $K-1$ users and the $K$th receiver are not greater than $1$ (the direct link gain), or (ii) the first $K-1$ users have the same average power constraint and the same channel gains for the links to the $K$th receiver. For other cases, the bounds match in terms of the secure degree of freedom, which is a high SNR characterization of the secrecy sum rate, but the gap between the upper bound and the lower bound we derived is in general a function of both $K$ and the channel gains.

The technique to derive the upper bound on the sum rate entails dividing the many-to-one channel into a point-to-point link from the $K$th user to the $K$th receiver, and a multiple access wiretap channel with an orthogonal main channel composed of the first $K-1$ users and their receivers [7], and upper bounding the secrecy sum rate of the multiple access wiretap channel part [8].

To derive the achievable rate, we utilize structured codes and layered coding [11], [15]. The design of the layers follows [15], but for each layer, though notably, we use nested lattice codes from [18] instead of the sphere shaped lattice codes as in [15]. The merit of using nested lattice codes for secrecy problems has recently been demonstrated in [11]; and the equivocation in this case is computable utilizing the nested structure.

We must also note that combining layered coding and nested lattice codes has been used in [11] and [13] to prove achievability results for the fully connected interference channel with confidential messages. As shown in [11] and [13], in general, the power allocated to each layer is limited in order for their interference to be "cancelable" at multiple receivers. This in essence forces us to increase the number of layers to support larger average power constraints. Unfortunately, increasing the number of layers leads to increased rate at which the information leaks to the eavesdropper. Luckily, for the many-to-one channel, since the channel is only sparsely connected, the number of layers is bounded. In fact, it is only a function of the number of users $K$ [15]. This means, if each layer leaks information at a certain rate, the total rate at which the information leaked to the eavesdropper is just a function of $K$ and does not grow unbounded with the average power constraint. This fact allows us to identify the two cases described earlier in the introduction where the gap between the lower and upper bound on the secrecy sum rate is only a function of the number of users $K$.

The remainder of the paper is organized as follows: In Section II, we describe the channel model. Section III states the main results, i.e., the upper bound and lower bound on the secrecy sum rate. Section IV proves the lower bound. Section V proves the upper bound. Section VI concludes the paper.

## II. THE GAUSSIAN MANY-TO-ONE INTERFERENCE CHANNEL WITH CONFIDENTIAL MESSAGES

### A. Model

The channel model is shown in Figure 1. There are $K$ users, denoted by node $S_1, S_2, ..., S_K$. User $k$ transmits a message $W_k$ to its receiver $D_k$. The network is sparsely connected in the sense that only the $K$th user is experiencing interference from the other $K-1$ users. There is no interference among the first $K-1$ users. Let $X_i, Y_i$ denote the transmitted and received signals for user $i$. Let $\sqrt{a_i}$ denote the channel gain between user $i$ and receiver $K$. Then the channel can be described formally as:

$$Y_k = X_k + Z_k, \quad , 1 \leq k \leq K-1 \tag{1}$$

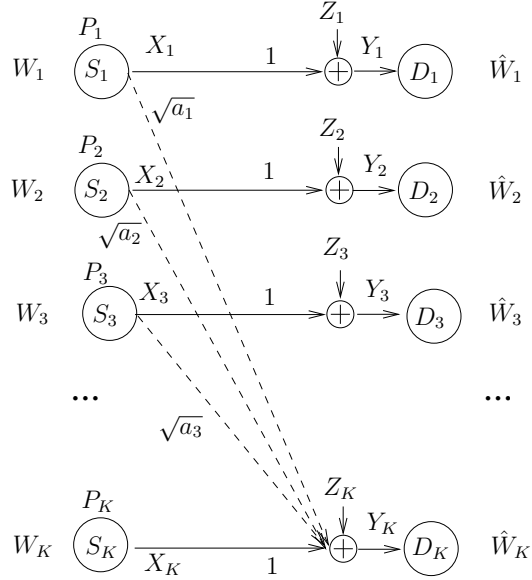

Fig. 1. Many-to-one Gaussian interference channel with $K$ users

$$Y_K = \sum_{i=1}^{K-1} \sqrt{a_i} X_i + Z_K \quad (2)$$

where $Z_k$s are independent zero mean Gaussian random variables with unit variance.

Let $n$ be the total number of channel uses. Since for $W_1, ..., W_{K-1}$, node $D_K$ is an eavesdropper, which receives signal $Y_K^n$, we have the following secrecy constraint:

$$\lim_{n \to \infty} \frac{1}{n} H(W_1, ..., W_{K-1} | Y_K^n) = \lim_{n \to \infty} \frac{1}{n} H(W_1, ..., W_{K-1}) \quad (3)$$

Let $\hat{W}_k$ be the decoding result of $W_k$ computed by node $D_k$. Then in order for $W_k$ to be transmitted reliably, we need:

$$\lim_{n \to \infty} \Pr(W_k \neq \hat{W}_k) = 0 \quad (4)$$

The average power constraint for node $S_k$ is $P_k$. Let $X_{k,i}$ denote the value of $X_k$ during the $i$th channel use. Then the power constraints can be expressed as:

$$\lim_{n \to \infty} \frac{1}{n} \sum_{i=1}^{n} E\left[X_{k,i}^2\right] \leq P_k \quad (5)$$

*Remark 1:* We restrict $X_i, Y_i$, $i = 1, ..., K$ to be real in this work. In [15], $X_i, Y_i$, $i = 1, ..., K$ are complex numbers. In general, real and complex interference channel with confidential messages can have very different achievable schemes; see [11] for example. However, for the many-to-one interference channel, whether the model is real or complex does not lead to any difference in terms of the results presented in this work. In a complex model, since the network is sparsely connected, all phases of the complex links can be canceled by pre-multiplication at the transmitters and post-multiplication at the receiver. The channel then becomes a real many-to-one interference channel. Therefore the same achievable scheme and converse derived in this work is applicable to the complex channel model as well. □

*B. Metric*

Define $R_k^s$ as the secrecy rate of the $k$th user:

$$R_k^s = \lim_{n\to\infty} \frac{1}{n} H(W_k), \quad 1 \leq k \leq K \tag{6}$$

The secrecy rate region is defined as all $(R_1^s, ....R_K^s)$ vectors such that the constraints (3), (4) and (5) are fulfilled.

The secrecy sum rate $R_\Sigma^s$ is simply

$$R_\Sigma^s = \sum_{k=1}^{K} R_i^s \tag{7}$$

The secure degree freedom of the secrecy sum rate is a high SNR characterization of $R_\Sigma^s$ and is defined as:

$$\limsup_{\substack{P_i=P, i=1,...,K \\ P\to\infty}} \frac{R_\Sigma^s}{\log_2(P)/2} \tag{8}$$

### III. MAIN RESULTS

In this section, we state the main results for the Gaussian many-to-one interference channel with confidential messages.

**1) *Secrecy Sum Rate:***

The lower bound and the upper bound on the secrecy sum rate are given in the next two theorems:

*Theorem 1:* Define $\bar{i}$ and $\tilde{i}$ as

$$\bar{i} = \arg\max_i \{a_i P_i, 1 \leq i \leq K-1\} \tag{9}$$

$$\tilde{i} = \arg\min_i \{a_i, 1 \leq i \leq K-1\} \tag{10}$$

The secrecy sum rate is lower bounded by:

$$\sum_{k=1}^{K} \max\{0, \frac{1}{2} \log_2 (P_k)\} - \max\{0, \frac{1}{2} \log_2 \left(\frac{a_{\bar{i}} P_{\bar{i}}}{\max\{1, a_{\bar{i}}\}}\right)\} - f(K) \tag{11}$$

where

$$f(K) = (2K - 1) \left(\frac{K-1}{2} + \frac{K+1}{2} \log_2 (K)\right) + \frac{1}{2} \log_2 (K) \tag{12}$$

*Theorem 2:* The secrecy sum rate is upper bounded by

$$R_{\Sigma}^{s} \leq \sum_{i=1}^{K} C(P_i) - C \left(\frac{\sum_{i=1}^{K-1} a_i P_i}{(K-1)c}\right) \tag{13}$$

where $C(x) = \frac{1}{2} \log_2(1 + x)$. $c = \max\{1, a_i, i = 1...K - 1\}$.

The derivation of the achievable rate in Theorem 1 is given in Section IV. The upper bound in Theorem 2 is derived in Section V.

### 2) *Secure Degree of Freedom:*

From Theorem 2 and Theorem 1, we observe the secure degree of freedom for the secrecy sum rate defined in (8) is equal to $K-1$ if there is at least one $i$ such that $a_i \neq 0, 1 \leq i \leq K-1$. Note that for the many-to-one interference channel without secrecy constraints, the degree of freedom for the sum rate is also $K - 1$ [15], which was shown by giving the transmitted signals of users $2, ..., K - 1$ to user $K$ as genie information. *Hence imposing secrecy constraints does not lead to any loss in terms of the degree of freedom for the sum rate.*

### 3) *Constant Gap Results:*

Secure degree of freedom provides the characterization of the secrecy sum rate at high SNR. For finite SNR, the gap between the lower bound from Theorem 1 and the upper bound from Theorem 2 is, in general, a function of both user number $K$ and the channel gains $\sqrt{a_i}$. However, in two cases shown below, we find this gap to be unaffected by the channel gains.

*Corollary 1:* When either of the following two conditions holds, the gap between the upper bound and the lower bound on the secrecy sum rate given by Theorem 2 and Theorem 1 is bounded by a constant which is only a function of $K$.

1) When all interference links are weaker than direct links: $0 \leq a_i < 1$, $1 \leq i \leq K - 1$, or
2) Symmetric SNR: $a_i = a, P_i = P, 1 \leq i \leq K - 1$

We provide the proof, which is immediate, next:

*Proof:* When $0 \leq a_i < 1$, $1 \leq i \leq K-1$, we have $c = 1$ in Theorem 2. Hence the upper bound is

$$\sum_{k=1}^{K} C(P_k) - \frac{1}{2}\log_2\left(K-1+\sum_{i=1}^{K-1} a_i P_i\right) + \frac{1}{2}\log_2(K-1) \tag{14}$$

On the other hand, the lower bound is lower bounded by

$$\sum_{k=1}^{K} \max\{0, \frac{1}{2}\log_2(P_k)\} - \frac{1}{2}\log_2\left(K-1+\sum_{i=1}^{K-1} a_i P_i\right) - f(K) \tag{15}$$

Since

$$\frac{1}{2}\log_2(1+x) - \max\left\{0, \frac{1}{2}\log_2 x\right\} \tag{16}$$

$$= \min\left\{\frac{1}{2}\log_2(1+x), \frac{1}{2}\log_2\left(\frac{1+x}{x}\right)\right\} \leq \frac{1}{2} \tag{17}$$

we find the difference between (15) and (14) is bounded by

$$\frac{K}{2} + \log_2(K-1) + f(K) \tag{18}$$

which is only a function of $K$.

When $a_i = a$, $P_i = P$, $1 \leq i \leq K-1$, the upper bound is

$$\sum_{k=1}^{K} C(P_k) - \frac{1}{2}\log_2(1+\min\{1,a\}P) \tag{19}$$

On the other hand, the lower bound is lower bounded by

$$\sum_{k=1}^{K} \max\{0, \frac{1}{2}\log_2(P_k)\} - \frac{1}{2}\log_2(1+\min\{1,a\}P) - f(K) \tag{20}$$

we find the difference between (20) and (19) is bounded by

$$\frac{K}{2} + f(K) \tag{21}$$

which is again only a function of $K$.

∎

## IV. ACHIEVABLE SCHEME

*A. Some Useful Results on Nested Lattice Codes*

In this section, we state some results on decoding nested lattice codes from [11] for the reader's convenience. The proofs of these results can be found in [11].

We begin by introducing some notations. A nested lattice code is defined as an intersection of an $N$-dimensional "fine" lattice $\Lambda$ and the fundamental region of an $N$-dimensional "coarse" lattice $\Lambda_c$, denoted by $\mathcal{V}(\Lambda_c)$. The modulus operation is defined as the quantization error of a point $x$ with respect to the coarse lattice $\Lambda_c$:

$$x \mod \Lambda_c = x - \arg\min_{u \in \Lambda_c} \|x - u\|_2 \tag{22}$$

where $\|x - y\|_2$ is the Euclidean distance between $x$ and $y$ in $\mathbf{R}^N$.

The signal $X^N$ transmitted over $N$ channel uses from a nested lattice codebook is given by

$$X^N = (u^N + d^N) \mod \Lambda_c \tag{23}$$

Here $u^N$ is the lattice point chosen from $\Lambda \cap \mathcal{V}(\Lambda_c)$, and $d^N$ is called the dithering vector. Conventionally, $d^N$ is defined as a continuous random vector which is uniformly distributed over $\mathcal{V}(\Lambda_c)$ [18]. As shown in [11], a fixed dithering vector can be used instead. Either way, the nature of $d^N$ will not affect the result described below. In the following, we assume $u^N$ is independent from $d^N$. We also assume that any dithering vector is perfectly known by all receiving nodes, the point being that it cannot be used to enhance secrecy.

Due to the employment of a uniformly distributed dithering vector $d^N$, the average power of a nested lattice codebook $\Lambda \cap \mathcal{V}(\Lambda_c)$ per channel use can be computed as:

$$P = \frac{1}{N \, \mathbf{vol}(\mathcal{V}(\Lambda_c))} \int_{x \in \mathcal{V}(\Lambda_c)} \|x\|_2^2 \, dx \tag{24}$$

where $\mathbf{vol}(\mathcal{V}(\Lambda_c))$ is the volume of the set $\mathcal{V}(\Lambda_c)$.

The rate $R$ of a nested lattice code book $\Lambda \cap \mathcal{V}(\Lambda_c)$ is given by

$$R = \frac{1}{N} \log_2 |\mathcal{V}(\Lambda_c) \cap \Lambda| \tag{25}$$

where $|S|$ is the cardinality of a set $S$.

We next describe a result on decoding nested lattice codes when interference is present.

Consider $K+1$ $N$-dimensional lattices $\Lambda_{p,i}, i = 0, ..., K$, such that $\Lambda_{p,i} \subset \mathbf{R}^N, i = 0, ..., K$ is Rogers-good for covering and Poltyrev-good for channel coding [19].

Construct the fine lattice $\Lambda$ as in [18, Section 7] such that $\Lambda_{p,0} \subset \Lambda$. Hence $\{\Lambda, \Lambda_{p,0}\}$ forms a nested lattice pair.

Define independent random variables $U_0^N, U_1^N, ...U_K^N$, such that $U_i^N, i = 0, ..., K$ is uniformly distributed over the fundamental region of $\Lambda_{p,i}$.

Define $\sigma^2(U_i), i = 0, ..., K$ as the variance per dimension of $U_i^N$. When $N$ increases, we scale $\Lambda$ and $\Lambda_{p,i}, i = 0, ..., K$ such that $\sigma^2(U_i)$ remains unchanged.

Define $Z^N$ as a $N$-dimensional vector which is composed of zero mean i.i.d. Gaussian random variables, each with variance $\sigma^2$. $Z^N$ is independent from $U_i^N, i = 1, ..., K$.

Define $t^N$ as a lattice point in $\Lambda \cap \mathcal{V}(\Lambda_{p,0})$. $t^N$ is independent from $Z^N$ and $U_i^N, i = 1, ..., K$.

Define the notation $\sum i = 1^j a_i = 0$ if $j < i$.

Define $Y^N$ as

$$Y^N = \left(t^N + \sum_{i=1}^{K} U_i^N + Z^N\right) \bmod \Lambda_{p,0} \tag{26}$$

Define $\tilde{t}^N$ as the value for $t^N$ decoded from $Y^N$ using an Euclidean distance decoder:

$$\tilde{t}^N = \arg \min_{u^N \in \Lambda \cap \mathcal{V}(\Lambda_{p,0})} \left\|Y^N - u^N\right\|_2 \tag{27}$$

Let $R_0$ be the rate of the nested lattice code book $\Lambda \cap \mathcal{V}(\Lambda_{p,0})$.

With these notations, we have the following results:

*Lemma 1:* [11] Define $\sum_{i=1}^{0} \sigma^2(U_i) = 0$. If

$$R_0 < \frac{1}{2} \log_2 \left(\frac{\sigma^2(U_0)}{\sigma^2 + \sum_{i=1}^{K} \sigma^2(U_i)}\right) \tag{28}$$

then for each $N$ dimension there exist lattices $\Lambda, \Lambda_{p,i}, t = 0, ...K$ such that $\Pr(t^N \neq \tilde{t}^N)$ decreases exponentially fast with $N$.

The following result is adapted from [18, (89)] and can be found in [11].

*Lemma 2:* Define $\mu$ as

$$\mu = \frac{\sigma^2(U_0)}{\sigma^2 + \sum_{i=1}^{K} \sigma^2(U_i)} \tag{29}$$

Then if $\mu > 1$, the probability

$$\Pr\left(\sum_{i=1}^{K} U_i^N + Z^N \bmod \Lambda_{p,0} \neq \sum_{i=1}^{K} U_i^N + Z^N\right) \tag{30}$$

decreases exponentially fast with respect to $N$.

To derive the secrecy rate, we need the following result from [11].

*Theorem 3:* [11] Let $t_1, t_2, ..., t_K$ be $K$ numbers taken from the fundamental region of a given lattice $\Lambda$. There exists a integer $T$, such that $1 \leq T \leq K^N$, and $\sum_{k=1}^{K} t_k$ is uniquely determined by $\{T, \sum_{k=1}^{K} t_k \mod \Lambda\}$.

Theorem 3 was used in [11] to bound the rate of information leaked to the eavesdropper. In this work, bounding this rate can be formulated as the following problem:

Consider a nested lattice pair $(\Lambda_f, \Lambda_c)$, $\Lambda_c \subset \Lambda_f$. Let $X_k^N = (t_k^N + d_k^N), k = 1, ..., K$, where $t_k^N \in \Lambda_f \cap \mathcal{V}(\Lambda_c)$. $t_k^N, d_k^N, k = 1, 2, ...K$ are independent. $t_K^N$ is uniformly distributed over $\Lambda_f \cap \mathcal{V}(\Lambda_c)$. Then we need to find a upper bound on

$$I\left(t_1^N, ..., t_{K-1}^N; \sum_{k=1}^{K} X_k^N, d_k^N, k = 1, ..., K\right) \tag{31}$$

to measure the rate of information leaked to the eavesdropper. The reason for this will become apparent in Section IV-B2.

*Lemma 3:*

$$I\left(t_1^N, ..., t_{K-1}^N; \sum_{k=1}^{K} X_k^N, d_k^N, k = 1, ..., K\right) \leq N \log_2 K \tag{32}$$

*Proof:* We use the notation $d^N$ to denote $d_k^N, k = 1, ..., K$. From Theorem 3, (31) can be written as:

$$I\left(t_1^N, ..., t_{K-1}^N; \sum_{k=1}^{K} X_k^N \mod \Lambda_c, d^N, T\right) \tag{33}$$

where $T$ is the integer in Theorem 3. $1 \leq T \leq K^N$.

$$I\left(t_1^N, ..., t_{K-1}^N; \sum_{k=1}^{K} X_k^N, d^N\right) \tag{34}$$

$$= I\left(t_1^N, ..., t_{K-1}^N; \sum_{k=1}^{K} X_k^N \mod \Lambda_c, d^N, T\right) \tag{35}$$

$$= I\left(t_1^N, ..., t_{K-1}^N; \sum_{k=1}^{K} (t_k^N + d^N) \mod \Lambda_c, d^N, T\right) \tag{36}$$

$$= I\left(t_1^N, ..., t_{K-1}^N; \sum_{k=1}^{K} t_k^N \mod \Lambda_c, d^N, T\right) \tag{37}$$

$$\leq I\left(t_1^N, ..., t_{K-1}^N; \sum_{k=1}^{K} t_k^N \bmod \Lambda_c, d^N\right) + H(T) \tag{38}$$

$$= I\left(t_1^N, ..., t_{K-1}^N; \sum_{k=1}^{K} t_k^N \bmod \Lambda_c\right) + H(T) \tag{39}$$

We next use the fact that $\Lambda_f \cap \mathcal{V}(\Lambda_c)$ is a Abelian group with the operation $x + y \bmod \Lambda_c$. Using it along with the fact that $t_K^N$ is uniformly distributed over $\Lambda_f \cap \mathcal{V}(\Lambda_c)$ and is independent from $t_1^N, ..., t_{K-1}^N$, we have

$$I\left(t_1^N, ..., t_{K-1}^N; \sum_{k=1}^{K} t_k^N \bmod \Lambda_c\right) = 0 \tag{40}$$

Hence (39) is upper bounded by $H(T) \leq N \log_2 K$ and we have proved the lemma. ∎

## B. Layered Coding Scheme

To prove the achievable rates we use a layered coding scheme [15]. This means the signal transmitted by each node will be the sum of the signals assigned for each layer. Let $X_k^N$ denote the signals transmitted by the $k$th user at the $m$th layer over $N$ channel uses. Let $X_{k,m}^N$ denote the term in $X_k^N$ that comes from the $m$th layer. Let the total number of layers be $M + 1$. Then layered coding leads to

$$X_k^N = \sum_{m=0}^{M} X_{k,m}^N \tag{41}$$

$X_{k,m}^N$ is just an all zero vector if no power is assigned to the $m$th layer by the $k$th user.

The signals received by the $k$th receiver $1 \leq k < K$ over $N$ channel uses is simply

$$Y_k^N = \sum_{m=0}^{M} X_{k,m}^N + Z_k^N \tag{42}$$

Hence the signal component received by each of the first $K - 1$ receivers is naturally expressed as the sum of signals from at most $M + 1$ layers. If a codebook with a proper average power and rate is chosen for each layer, the receiver can process the layers sequentially. This means, when processing the $m$th layer, the receiver decodes $X_{k,m}^N$, subtracts it from $Y_k^N$, and uses the result as the input when processing the $m - 1$th layer.

For the $K$th receiver, its received signal over $N$ channel uses is given by

$$Y_K^N = \sum_{k=1}^{K-1} \sqrt{a_k} X_k^N + X_K^N + Z_K^N \tag{43}$$

$$= \sum_{k=1}^{K-1} (\sqrt{a_k} \sum_{m=0}^{M} X_{k,m}^N) + \sum_{m=0}^{M} X_{K,m}^N + Z_K^N \quad (44)$$

$$= \sum_{m=0}^{M} (X_{K,m}^N + \sum_{k=1}^{K-1} \sqrt{a_k} X_{k,m}^N) + Z_K^N \quad (45)$$

Again, if a codebook with a proper average power and rate is chosen for the each layer, the receiver can subtract the influence of $X_{K,m}^N + \sum_{k=1}^{K-1} \sqrt{a_k} X_{k,m}^N$ after processing layer $m$ and move on to layer $m-1$.

The maximal power that $X_{k,m}^N$ can use is determined in the same way as [15]. As shown in [15], we first determine the so-called delimiter of each layer based on the following procedure:

1) Compute the set $\tilde{Q}_1$ as the union of the pairs $\{a_i P_i, a_i\}, i = 1, ..., K-1$.
2) Let $\tilde{Q}_2$ be the subset of $\tilde{Q}_1$ which only contains numbers that are greater than 1.
3) Let $\tilde{Q}$ be the union of $\tilde{Q}_2$ and the pair $\{P_K, 1\}$.
4) Compute the list $Q$ such that the $i$th number in $Q$ is the $i$th smallest number in the set $\tilde{Q}$, $i \geq 0$.
5) The number of layers $M+1$ equals the cardinality of the list $Q$.
6) The delimiter of the 0th layer is defined as $(-\infty, 1]$. The delimiter of the $m$th layer, $m \geq 1$, is given by $[q_{m-1}, q_m]$, where $q_i$ is the $i$th number in the list $Q$.

Let $P_{k,m}$ be the maximal power that the $k$th user can spend on the $m$th layer, i.e., $E[X_{k,m}^2] \leq P_{k,m}$. Then $P_{k,m}$ is computed from the delimiters of the layers as follows [15]:

1) For the first $K-1$ users, the power allocated to each layer is computed as follows: For the $i$th user, $1 \leq k \leq K-1$,
   a) At most $P_{k,0} = \max\{1/a_i - 1, 0\}$ power is allocated to the 0th layer.
   b) If $a_i P_i \geq q_{m+1}$ and $q_m \geq a_i$, at most $(q_{m+1} - q_m)/a_i$ is allocated to the $m$th layer. Otherwise, no power is allocated to the $m$th layer.
2) The power allocation of the $K$th user among different layers is determined as follows:
   a) No power is allocated to the 0th layer.
   b) If $P_K \geq q_{m+1}$, at most $q_{m+1} - q_m$ power is allocated to the $m$th layer. Otherwise, no power is allocated to the $m$th layer.

Note that, with this power allocation among layers, it can be verified that the total power of the $i$th user does not exceed $P_i$.

We next describe the codebook used at each layer: We use $\mathcal{C}_{k,m}$ to denote the codebook used for the $m$th layer by the $k$th user. If the maximal power this user can allocate to the $m$th layer is 0, then $\mathcal{C}_{k,m}$ only contains an all zero vector. Otherwise, we construct $\mathcal{C}_{k,m}$ as follows: We start from a $N$-dimensional nested lattice codebook denoted by $(\Lambda_{f,m}, \Lambda_{c,m})$, $\Lambda_{c,m} \subset \Lambda_{f,m}$, with average power 1 and rate $R_m$. Then, we scale every codeword in the codebook with $\sqrt{P_{k,m}}$ so it has average power $P_{k,m}$ and let it be $\mathcal{C}_{k,m}$. This means for a given layer $m$, $\mathcal{C}_{k,m}, k = 1, 2, ..., K$ are scaled forms of the same nested lattice codebook $(\Lambda_{f,m}, \Lambda_{c,m})$. Hence, $X^N_{k,m}$ is given by

$$X^N_{k,m} = \sqrt{P_{k,m}}(t^N_{k,m} + d^N_{k,m}) \bmod \Lambda_{c,m}, \quad m > 0 \tag{46}$$

where $t^N_{k,m} \in \Lambda_{f,m} \cap \mathcal{V}(\Lambda_{c,m})$. $d^N_{k,m}$ is the dither vector.

To simplify the expressions, we set $a_K = 1$. Let $\mathcal{U}_m$ denote the set of users that can allocate nonzero power to the $m$th layer. Then, for $m > 0$, we can write

$$X^N_{K,m} + \sum_{k=1}^{K-1} \sqrt{a_k} X^N_{k,m} \tag{47}$$

$$= \sum_{k \in \mathcal{U}_m} \sqrt{a_k P_{k,m}} \left(t^N_{k,m} + d^N_{k,m}\right) \bmod \Lambda_{c,m} \tag{48}$$

Notice that the power allocation among layers are chosen such that $a_k P_{k,m}$ equals $q_{m+1} - q_m$. Hence (48) equals:

$$\sqrt{q_{m+1} - q_m} \sum_{k \in \mathcal{U}_m} \left(t^N_{k,m} + d^N_{k,m}\right) \bmod \Lambda_{c,m} \tag{49}$$

*1) Decoding:* We next describe the decoding process at each receiver [11]. First we describe the decoding process at receiver $K$. The decoder starts from layer $M$ and processes the layers according to their index in decreasing order. It subtracts the influence of the signals from a layer after it is decoded. Since user $K$ does not transmit at the 0th layer, receiver $K$ processes $M$ layers in all.

Define $Y^N_{K,\bar{m}}$ as the partial sum in $Y^N_K$:

$$Y^N_{K,\bar{m}} = \sum_{m=0}^{\bar{m}} (X^N_{K,m} + \sum_{k=1}^{K-1} \sqrt{a_k} X^N_{k,m}) + Z^N_K \tag{50}$$

When processing the $\bar{m}$th layer, receiver $K$ first computes:

$$(Y^N_{K,\bar{m}} - \sqrt{q_{\bar{m}+1} - q_{\bar{m}}} \sum_{k \in \mathcal{U}_{\bar{m}}} d^N_{k,\bar{m}}) \bmod \sqrt{q_{\bar{m}+1} - q_{\bar{m}}} \Lambda_{c,\bar{m}} \tag{51}$$

Applying (49), we find (51) equals:

$$(\sqrt{q_{\bar{m}+1} - q_{\bar{m}}} \sum_{k \in \mathcal{U}_{\bar{m}}} t^N_{k,\bar{m}} + \sum_{m=0}^{\bar{m}-1}\left(X^N_{K,m} + \sum_{k=1}^{K-1}\sqrt{a_k}X^N_{k,m}\right) + Z^N_K) \bmod \sqrt{q_{\bar{m}+1} - q_{\bar{m}}}\Lambda_{c,\bar{m}} \quad (52)$$

Note (52) has the same form as (26), where $\sqrt{q_{\bar{m}+1} - q_{\bar{m}}}\Lambda_{c,\bar{m}}$ corresponds to $\Lambda_{p,0}$. $\sqrt{q_{\bar{m}+1} - q_{\bar{m}}} \sum_{k \in \mathcal{U}_{\bar{m}}} t^N_{k,\bar{m}} \bmod \sqrt{q_{\bar{m}+1} - q_{\bar{m}}}\Lambda_{c,\bar{m}}$ corresponds to $t^N$ in (26). Then, according to Lemma 1, if

$$0 \leq R_{\bar{m}} < \max\{0, \frac{1}{2}\log_2(\frac{q_{\bar{m}+1} - q_{\bar{m}}}{1 + \sum_{m=0}^{\bar{m}-1}\sum_{k \in \mathcal{U}_m} a_k P_{k,m}})\} \quad (53)$$

then receiver $K$ can decode $(\sqrt{q_{\bar{m}+1} - q_{\bar{m}}} \sum_{k \in \mathcal{U}_{\bar{m}}} t^N_{k,\bar{m}}) \bmod \sqrt{q_{\bar{m}+1} - q_{\bar{m}}}\Lambda_{c,\bar{m}}$ from (52) with high probability. We next derive a lower bound for the right hand side of (53) like [15]. From the power allocation among layers, we observe

$$a_k P_{k,0} \leq 1, \quad 1 \leq k \leq K - 1 \quad (54)$$

and for $m \geq 1$,

$$a_k P_{k,m} \leq q_{m+1} - q_m \quad (55)$$

$$|\mathcal{U}_m| \leq K \quad (56)$$

Applying these results to the right hand side of (53), we find it is lower bounded by:

$$\max\{0, \frac{1}{2}\log_2 \frac{q_{\bar{m}+1} - q_{\bar{m}}}{1 + \sum_{m=1}^{\bar{m}-1}(K(q_{m+1} - q_m)) + (K-1)}\} \quad (57)$$

$$= \max\{0, \frac{1}{2}\log_2 \frac{q_{\bar{m}+1} - q_{\bar{m}}}{Kq_{\bar{m}}}\} \quad (58)$$

Hence a sufficient condition for receiver $K$ to be able to decode $(\sqrt{q_{\bar{m}+1} - q_{\bar{m}}} \sum_{k \in \mathcal{U}_{\bar{m}}} t^N_{k,\bar{m}}) \bmod \sqrt{q_{\bar{m}+1} - q_{\bar{m}}}\Lambda_{c,\bar{m}}$ is to require

$$0 < R_{\bar{m}} < \max\{0, \frac{1}{2}\log_2 \frac{q_{\bar{m}+1} - q_{\bar{m}}}{Kq_{\bar{m}}}\} \quad (59)$$

After decoding $(\sqrt{q_{\bar{m}+1} - q_{\bar{m}}} \sum_{k \in \mathcal{U}_{\bar{m}}} t^N_{k,\bar{m}}) \bmod \sqrt{q_{\bar{m}+1} - q_{\bar{m}}}\Lambda_{c,\bar{m}}$, receiver $K$ subtracts it from (52) and obtain

$$(\sum_{m=0}^{\bar{m}-1}\left(X^N_{K,m} + \sum_{k=1}^{K-1}\sqrt{a_k}X^N_{k,m}\right) + Z^N_K) \bmod \sqrt{q_{\bar{m}+1} - q_{\bar{m}}}\Lambda_{c,\bar{m}} \quad (60)$$

From (59), we notice if $R_{\bar{m}} > 0$, we must have

$$q_{\bar{m}+1} - q_{\bar{m}} > Kq_{\bar{m}} \quad (61)$$

Note that the right hand side of (61) is an upper bound on the average power per channel use of the term inside the modulus operation in (60). Hence, we can apply Lemma 2 and find that (60) equals

$$\sum_{m=0}^{\bar{m}-1} \left( X_{K,m}^N + \sum_{k=1}^{K-1} \sqrt{a_k} X_{k,m}^N \right) + Z_K^N \tag{62}$$

with high probability. Thus, receiver $K$ can use (60) when it decodes layer $m-1$.

We next describe the decoding procedure at receiver $k$, $1 \leq k \leq K-1$. Receiver $k$ processes layers in a way similar to receiver $K$, except that it starts from layer $m$ such that $q_{m+1} = a_i P_i$. If $a_i \geq 1$, the index of the last layer it processes is the one whose $q_m = a_i$. Otherwise, it decodes all the way to the 0th layer.

We define $Y_{k,\bar{m}}^N$ as the partial sum in $Y_k^N$:

$$Y_{k,\bar{m}}^N = \sum_{m=0}^{\bar{m}} X_{k,m}^N + Z_k^N \tag{63}$$

When processing the $\bar{m}$th layer, receiver $k$ first computes:

$$(Y_{k,\bar{m}} - d_{k,\bar{m}}^N) \bmod \Lambda_{c,\bar{m}} \tag{64}$$

$$= (t_{k,\bar{m}}^N + \sum_{m=0}^{\bar{m}-1} X_{k,m}^N + Z_k^N) \bmod \Lambda_{c,\bar{m}} \tag{65}$$

Again since (65) has the same form as (26), according to Lemma 1, if

$$0 < R_{\bar{m}} < \max \left\{ 0, \frac{1}{2} \log_2 \frac{P_{k,\bar{m}}}{\sum_{m=0}^{\bar{m}-1} P_{k,m} + 1} \right\} \tag{66}$$

then receiver $k$ can decode $t_{k,\bar{m}}^N$ with high probability.

If $a_k \leq 1$, $P_{k,m} = (q_{m+1} - q_m)/a_k$, $m \geq 1$. $P_{k,0} = (1 - a_k)/a_k$. Hence the right hand side of (66) equals:

$$\max \left\{ 0, \frac{1}{2} \log_2 \frac{q_{\bar{m}+1} - q_{\bar{m}}}{a_k \left( \sum_{m=1}^{\bar{m}-1} \frac{q_{m+1} - q_m}{a_k} + \frac{1-a_k}{a_k} + 1 \right)} \right\} \tag{67}$$

$$= \max \left\{ 0, \frac{1}{2} \log_2 \frac{q_{\bar{m}+1} - q_{\bar{m}}}{q_{\bar{m}}} \right\} \tag{68}$$

If $a_k \geq 1$, there must exists $m_0$ such that $q_{m_0} = a_k$. The right hand side of (66) equals:

$$\max \left\{ 0, \frac{1}{2} \log_2 \frac{q_{\bar{m}+1} - q_{\bar{m}}}{a_k \left( \sum_{m=m_0}^{\bar{m}-1} \frac{q_m - q_{m-1}}{a_k} + 1 \right)} \right\} \tag{69}$$

$$= \max\left\{0, \frac{1}{2}\log_2 \frac{q_{\bar{m}+1} - q_{\bar{m}}}{q_{\bar{m}}}\right\} \tag{70}$$

Hence, in either case, we arrive at the following sufficient condition for receiver $k$ to decode $t_{k,\bar{m}}^N$:

$$0 \leq R_{\bar{m}} < \max\{0, \frac{1}{2}\log_2 \frac{q_{\bar{m}+1} - q_{\bar{m}}}{q_{\bar{m}}}\} \tag{71}$$

We observe (71) is a less stringent condition than (59).

After decoding $t_{k,\bar{m}}^N$, receiver $k$ subtracts it from (65) and obtain

$$(\sum_{m=0}^{\bar{m}-1} X_{k,m}^N + Z_k^N) \bmod \Lambda_{c,\bar{m}} \tag{72}$$

From (71), we notice if $R_{\bar{m}} > 0$, we must have

$$q_{\bar{m}+1} - q_{\bar{m}} > q_{\bar{m}} \tag{73}$$

Again the right hand side of (73) is an upper bound on the average power per channel use of the term inside the modulus operation in (72). Hence we can apply Lemma 2 and find (72) equals

$$\sum_{m=0}^{\bar{m}-1} X_{k,m}^N + Z_k^N \tag{74}$$

with high probability. Hence, receiver $k$ can use (72) when it decodes layer $m-1$.

With this, we have shown that the receiver $k$, $1 \leq k \leq K-1$, can decode $\{t_{k,m}^N : P_{k,m} > 0, m = 0, 1, ..., M\}$ and the receiver $K$ can decode $\{(\sum_{k \in \mathcal{U}_m} t_{k,m}^N) \bmod \Lambda_{c,m}, m = 0, 1, ..., M\}$ if (59) holds.

*2) Secrecy Rate Computation:* We next derive the achievable secrecy rate region for each layer, from which we will derive the achievable secrecy rate region for the whole channel. Define $\mathcal{U}_m'$ as:

$$\mathcal{U}_m' = \mathcal{U}_m \cap \{1, ..., K-1\} \tag{75}$$

We model the $m$th layer as the following equivalent channel, which takes lattice points $t_{k,m}^N$ as inputs, and produces $X_{k,m}^N$ as outputs to the $k$th receiver, $1 \leq k \leq K-1$ and $X_{\Sigma,m}^N$ as outputs to the eavesdropper (the $K$th receiver).

$$X_{k,m}^N = \left(t_{k,m}^N + d_{k,m}^N\right) \bmod \Lambda_{c,m}, \quad k \in \mathcal{U}_m'$$

$$X_{\Sigma,m}^N = \begin{cases} \sum_{k \in \mathcal{U}_m} \sqrt{a_k} X_{k,m}^N, & m \geq 1 \\ \sum_{k \in \mathcal{U}_m} \sqrt{a_k} X_{k,m}^N + Z_K^N & m = 0 \end{cases} \tag{76}$$

To achieve a trade-off between the secrecy rates of the lattice points between the first $K-1$ users and the $K$th user, for a layer $m$ such that $K \in \mathcal{U}_m$, two different modes of operation are possible:

1) The first $K-1$ users may choose to set $t_{k,m}^N = 0, k \in \mathcal{U}_m \cap \{1, 2, ..., K-1\}$ so that for layer $m$, receiver $K$ decodes $(\sum_{k \in \mathcal{U}_m} t_{k,m}^N) \mod \Lambda_{c,m} = t_{K,m}^N$.
2) The $K$th user may send $t_{K,m}^N$ by choosing it randomly from the nested lattice codebook. In this case, $t_{K,m}^N$ does not carry any information. Its purpose is to reduce the amount of information leaked to the $K$th receiver regarding the value of $t_{k,m}^N, k \in \mathcal{U}_m'$.

For the first mode, the achieved secrecy rate region is given by:

$$R_i = 0, \quad 1 \leq i \leq K-1 \tag{77}$$

$$0 \leq R_K \leq R_m \tag{78}$$

For the second mode, $R_K = 0$. We next find the achievable secrecy rate region for $R_i$, $1 \leq i \leq K-1$. Define $W_{k,m}$ as the confidential message that is transmitted over the $m$th layer by the $k$th user. Let the total number of channel uses be $n \times N$. Define $R_{k,m}^s$ as:

$$R_{k,m}^s = \lim_{n,N \to \infty} \frac{1}{nN} H(W_{k,m}) \tag{79}$$

For a set $S$, define $W_{S,m}$ and $t_{S,m}$ as:

$$W_{S,m} = \{W_{k,m} : k \in S\} \tag{80}$$

$$t_{S,m}^N = \{t_{k,m}^N : k \in S\} \tag{81}$$

Then we have

*Lemma 4:* Let $d_m^N$ be the shorthand for all dithering vectors used at layer $m$. The following secrecy rate region is achievable for the channel in (76):

$$0 \leq R_{k,m}^s \leq R_m - R_{k,m}^x, k \in \mathcal{U}_m' \tag{82}$$

$$\sum_{k \in \mathcal{U}_m'} R_{k,m}^x = \lim_{N \to \infty} \frac{1}{N} I\left(t_{\mathcal{U}_m',m}^N; X_{\Sigma,m}^N, d_m^N\right) \tag{83}$$

$$\sum_{k \in S} R_{k,m}^x \leq \lim_{N \to \infty} \frac{1}{N} I\left(t_{S,m}^N; X_{\Sigma,m}^N, d_m^N | t_{\mathcal{U}_m'/S,m}^N\right), \quad \forall S \subset \mathcal{U}_m' \tag{84}$$

$$0 \leq R_{k,m}^x \tag{85}$$

$$R_{k,m}^s = 0, k \in \{1, ..., K\}/\mathcal{U}_m' \tag{86}$$

The mutual information is evaluated with $t_{k,m}^N$ being independent and uniformly distributed over the $N$-dimensional nested lattice codebook.

*Proof:* The codebook of user $k$, $k \in \mathcal{U}_m'$, is a length-$nN$ real sequence sampled in an i.i.d. fashion uniformly from the nested lattice codebook $\mathcal{C}_{k,m}$. Each codebook is randomly binned into several bins, each containing $2^{nNR_{k,m}^x}$ codewords. Each bin is labeled from $1$ to $2^{nN(R_m - R_{k,m}^x)}$. User $k$ selects a codeword randomly from the bin indexed by $W_{k,m}$ and transmits it over $nN$ channel uses. Receiver $k$ can recover $t_{k,m}^N$. Hence, it can recover the length-$nN$ codeword and consequently, the bin index.

It remains to prove the secrecy constraint:

$$\lim_{N \to \infty} \frac{1}{N} I\left(W_{\mathcal{U}_m', m}; X_{\Sigma, m}^N, d_m^N\right) = 0 \tag{87}$$

holds. This can be shown as follows:

$$H\left(W_{\mathcal{U}_m', m} | X_{\Sigma, m}^{Nn}, d_m^{Nn}\right) \tag{88}$$

$$= H\left(W_{\mathcal{U}_m', m} | X_{\Sigma, m}^{Nn}, d_m^{Nn}\right) - H\left(W_{\mathcal{U}_m', m} | X_{\Sigma, m}^{Nn}, d_m^{Nn}, t_{\mathcal{U}_m', m}^{Nn}\right) \tag{89}$$

$$= I\left(W_{\mathcal{U}_m', m}; t_{\mathcal{U}_m', m}^{Nn} | X_{\Sigma, m}^{Nn}, d_m^{Nn}\right) \tag{90}$$

$$= H\left(t_{\mathcal{U}_m', m}^{Nn} | X_{\Sigma, m}^{Nn}, d_m^{Nn}\right) - H\left(t_{\mathcal{U}_m', m}^{Nn} | W_{\mathcal{U}_m', m}, X_{\Sigma, m}^{Nn}, d_m^{Nn}\right) \tag{91}$$

$$\geq H\left(t_{\mathcal{U}_m', m}^{Nn} | X_{\Sigma, m}^{Nn}, d_m^{Nn}\right) - n\varepsilon \tag{92}$$

$$= H\left(t_{\mathcal{U}_m', m}^{Nn}\right) - I\left(t_{\mathcal{U}_m', m}^{Nn}; X_{\Sigma, m}^{Nn}, d_m^{Nn}\right) - n\varepsilon \tag{93}$$

$$= nNR_{k,m} - I\left(t_{\mathcal{U}_m', m}^{Nn}; X_{\Sigma, m}^{Nn}, d_m^{Nn}\right) - n\varepsilon \tag{94}$$

$$\geq nNR_{k,m} - nI\left(t_{\mathcal{U}_m', m}^{N}; X_{\Sigma, m}^{N}\right) - n\varepsilon \tag{95}$$

Equation (90) follows because $W_{\mathcal{U}_m', m}$ is determined once $X_{\Sigma, m}^{Nn}, d_m^{Nn}$ is given. Equation (92) holds because given $W_{\mathcal{U}_m', m}$, the uncertainty in $t_{\mathcal{U}_m', m}^{Nn}$ can be resolved by receiver $K$ as its rates are within the capacity region of the multiple access channel with inputs $t_{\mathcal{U}_m', m}^{Nn}$ and output $X_{\Sigma, m}^{Nn}, d_m^{Nn}$. This is a consequence of (83) and (84). (95) follows because each $N$ channel uses are memoryless from the other $N$ channel uses.

Dividing both sides by $Nn$ and letting $N, n$ tend to $\infty$, we have (87). Hence we have proved the lemma. ∎

*Remark 2:* The secrecy sum rate of the region given by Lemma 4 is always lower bounded

by

$$|\mathcal{U}'_m|R_m - \lim_{N\to\infty} \frac{1}{N} I\left(t^N_{\mathcal{U}'_m,m}; X^N_{\Sigma,m}, d^N_m\right) \tag{96}$$

regardless of whether the region is empty or not. To show this, we can let $R^x_{k,m}$ equal $1/|\mathcal{U}'_m|$ times the right hand side of (83). Because the region defined by (83) and (84) is symmetric over $R^x_{k,m}$, the above choice of $R^x_{k,m}$ always fulfills (83) and (84). If $R^x_{k,m} > R_m$, then (96) is clearly a lower bound to the sum rate since it is negative. If $R^x_{k,m} < R_m$, then from (82) we observe (96) is a lower bound to the sum rate. □

Define $\mathcal{R}_m$, a region of $R^s_{k,m}$, as a convex hull of the region given by (82)-(86) and the region given by (77)-(78) if $m \geq 1$. $\mathcal{R}_0$ is just the region given by (82)-(86) since user $K$ does not transmit at layer 0. Then if $nN$ is the total number of channel uses, we have:

$$\lim_{nN\to\infty} \frac{1}{nN} I\left(W_{\mathcal{U}'_m,m}; X^{Nn}_{\Sigma,m}, d^{Nn}_m\right) = 0 \tag{97}$$

*Lemma 5:* The achievable secrecy rate region for the whole channel is given by

$$\sum_{m=0}^{M} \mathcal{R}_m \tag{98}$$

*Proof:* In each layer $m \geq 1$, we either let user $K$ transmit or let the first $K-1$ users perform the random binning scheme given in Lemma 4 and let the $K$th user send a randomly selected lattice code to protect the signals of the first $K-1$ users from leaking to receiver $K$. Using the sequential decoding and subtraction scheme described in Section IV-B1, the first $K-1$ receivers can recover the lattice points from their respective transmitter for each layer. The $K$th receiver can recover the lattice point transmitted by the $K$th user if the first $K-1$ users are not transmitting at this layer. From the sequences of recovered lattice points, each receiver can recover the index of the bin which contains these sequences and hence can recover the messages. Thus, clearly, the region (98) fulfills the requirement that messages must be transmitted reliably.

It remains to show that the messages of the first $K-1$ users are confidential from the $K$th receiver. Let $d^{Nn}$ be the shorthand of dithering vectors used at all layers. Then we can write:

$$I\left(\{W_{\mathcal{U}'_m,m} : m = 0...M\}; Y^N_K\right) \tag{99}$$

$$= I\left(\{W_{\mathcal{U}'_m,m} : m = 0...M\}; \sum_{m=0}^{M} X^{Nn}_{\Sigma,m}, d^{Nn}\right) \tag{100}$$

$$\leq I\left(\left\{W_{\mathcal{U}'_m,m}: m=0...M\right\}; \left\{X_{\Sigma,m}^{Nn}, d^{Nn}: m=0...M\right\}\right) \tag{101}$$

$$\leq \sum_{m=0}^{M} I\left(W_{\mathcal{U}_m,m}; X_{\Sigma,m}^{Nn}, d^{Nn}\right) \tag{102}$$

Since $\lim_{nN\to\infty} \frac{1}{nN} I\left(W_{\mathcal{U}'_m,m}; X_{\Sigma,m}^{Nn}, d^{Nn}\right) = 0$, we have

$$\lim_{nN\to\infty} \frac{1}{nN} I\left(\left\{W_{\mathcal{U}'_m,m}: m=0...M\right\}; Y_K^N\right) = 0 \tag{103}$$

Hence we have proved the lemma. ∎

Evaluating the region (98) is difficult due to the difficulty in deriving a lower bound to the right hand side of (84). However, a lower bound on the secrecy sum rate can still be found, which we describe below:

From the discussion in Remark 2, we notice a lower bound on the secrecy sum rate can be obtained by finding an upper bound on the right hand side of (83).

For $m \geq 1$, $I\left(t_{\mathcal{U}'_m,m}^N; X_{\Sigma,m}^N, d_m^N\right)$ can be upper bounded as follows:

$$I(t_{\mathcal{U}'_m,m}^N; X_{\Sigma,m}^N, d_m^N) \tag{104}$$

$$= I(t_{\mathcal{U}'_m,m}^N; X_{K,m}^N + \sum_{k \in \mathcal{U}'_m}^{K-1} \sqrt{a_k} X_{k,m}^N, d_m^N) \tag{105}$$

$$= I(t_{\mathcal{U}'_m,m}^N; \sqrt{q_{m+1}-q_m} \sum_{k \in \mathcal{U}_m} (t_{k,m}^N + d_{k,m}^N) \bmod \sqrt{q_{m+1}-q_m}\Lambda_{c,m}, d_m^N) \tag{106}$$

$$= I(t_{\mathcal{U}'_m,m}^N; \sum_{k \in \mathcal{U}_m} (t_{k,m}^N + d_{k,m}^N) \bmod \Lambda_{c,m}, d_m^N) \tag{107}$$

From Lemma 3, we find (107) is upper bounded by $N \log_2 |\mathcal{U}_m|$.

Hence

$$\lim_{N\to\infty} \frac{1}{N} I\left(t_{\mathcal{U}'_m,m}^N; X_{\Sigma,m}^N, d_m^N\right) \leq \log_2 |\mathcal{U}_m| \tag{108}$$

Let the secrecy sum rate provided by each layer $m$ be denoted by $R_{\Sigma,m}^s$. Hence we $R_{\Sigma,m}^s$, $m \geq 1$, is lower bounded by:

$$R_{\Sigma,m}^s \geq |\mathcal{U}'_m| R_m - \log_2 K \tag{109}$$

$$= \frac{|\mathcal{U}'_m|}{2} \max\left\{0, \log_2 \frac{q_{m+1}-q_m}{K q_m}\right\} - \log_2 K \tag{110}$$

For the 0th layer, since the signals are not aligned on a lattice, the method we use to lower bound the sum rate is different. Recall that $d_0^{Nn}$ denotes the dithering vectors used by those users transmitting at the 0th layer, which are known by all receivers. We first upper bound the right hand side of (83) as shown below:

$$I\left(t_{\mathcal{U}_0',0}^N; X_{\Sigma,0}^{Nn}, d_0^{Nn}\right) = I\left(t_{\mathcal{U}_0',0}^N; X_{\Sigma,0}^{Nn}|d_0^{Nn}\right) \tag{111}$$

$$\leq I\left(t_{\mathcal{U}_0',0}^N, d_0^{Nn}; X_{\Sigma,0}^{Nn}\right) \tag{112}$$

$$= I\left(t_{\mathcal{U}_0',0}^N, d_0^{Nn}; \sum_{k \in \mathcal{U}_0} \sqrt{a_k} X_{k,0}^N + Z_k^N\right) \tag{113}$$

$$\leq I\left(X_{\mathcal{U}_0',0}^N; \sum_{k \in \mathcal{U}_0'} \sqrt{a_k} X_{k,0}^N + Z_k^N\right) \tag{114}$$

$$\leq NC\left(\sum_{k \in \mathcal{U}_0'} E\left[a_k X_{k,0}^2\right]\right) \tag{115}$$

$$\leq NC\left(|\mathcal{U}_0'|\right) \tag{116}$$

$$= \frac{N}{2} \log_2(K) \tag{117}$$

where the notation $C(x)$, as defined before, is $\frac{1}{2}\log_2(1+x)$. Therefore, from Lemma 4, we have

$$R_{\Sigma,m=0}^s \geq \sum_{k \in \mathcal{U}_0'} R_{k,0} - \frac{1}{2}\log_2(K) \tag{118}$$

Using (110) and (118), we are now ready to prove Theorem 1.

*Proof of Theorem 1:* From (110), $R_{\Sigma,m}^s$, $m \geq 1$ is lower bounded by

$$R_{\Sigma,m}^s \geq \frac{|\mathcal{U}_m'|}{2} \max\left\{0, \log_2 \frac{q_{m+1} - q_m}{Kq_m}\right\} - \log_2 K \tag{119}$$

$$\geq \frac{|\mathcal{U}_m'|}{2} \max\left\{0, \log_2 \frac{q_{m+1} - q_m}{q_m}\right\} - \frac{K-1}{2}\log_2 K - \log_2 K \tag{120}$$

Since $\max\{0, \log_2 x\} \geq \max\{0, \log_2(1+x) - 1\}$ [15], we find (120) is lower bounded by:

$$\frac{|\mathcal{U}_m'|}{2} \max\left\{0, \log_2\left(1 + \frac{q_{m+1} - q_m}{q_m}\right) - 1\right\} - \frac{K+1}{2}\log_2 K \tag{121}$$

$$\geq \frac{|\mathcal{U}_m'|}{2} \max\left\{0, \log_2\left(\frac{q_{m+1}}{q_m}\right)\right\} - \frac{|\mathcal{U}_m'|}{2} - \frac{K+1}{2}\log_2 K \tag{122}$$

$$\geq \frac{|\mathcal{U}_m'|}{2} \max\left\{0, \log_2\left(\frac{q_{m+1}}{q_m}\right)\right\} - \frac{K-1}{2} - \frac{K+1}{2}\log_2 K \tag{123}$$

From (118), the secrecy sum rate provided by layer 0, $R_{\Sigma,0}^s$, is lower bounded by:

$$R_{\Sigma,0}^s \geq \sum_{k \in \mathcal{U}_0'} R_{k,0} - \frac{1}{2}\log_2(K) \tag{124}$$

$$\geq \sum_{k \in \mathcal{U}'_0} \frac{1}{2} \log_2 \left(\frac{1}{a_k}\right) - \frac{1}{2} \log_2 (K) \tag{125}$$

Define $f_M(K)$ as

$$f_M(K) = M\left(\frac{K-1}{2} + \frac{K+1}{2} \log_2(K)\right) + \frac{1}{2}\log_2(K) \tag{126}$$

Then we have

$$\sum_{m=0}^{M} R^s_{\Sigma,m} \geq \sum_{m=1}^{M} \frac{|\mathcal{U}'_m|}{2} \max\{0, \log_2\left(\frac{q_{m+1}}{q_m}\right)\} + \sum_{k \in \mathcal{U}'_0} \frac{1}{2} \log_2\left(\frac{1}{a_k}\right) - f_M(K) \tag{127}$$

Define $B_k$ as the

$$B_k = \{m : P_{k,m} > 0\} \tag{128}$$

Then (127) can be written as:

$$\sum_{k=1}^{K} \left(\sum_{m \in B_k} \max\{0, \frac{1}{2}\log_2\left(\frac{q_{m+1}}{q_m}\right)\} + \max\left\{\frac{1}{2}\log_2\left(\frac{1}{a_k}\right), 0\right\}\right) - f_M(K) \tag{129}$$

$$\geq \sum_{k=1}^{K} \max\{0, \frac{1}{2}\log_2(P_k)\} - \sum_{m \notin B_K} \max\{0, \frac{1}{2}\log_2\left(\frac{q_{m+1}}{q_m}\right)\} - f_M(K) \tag{130}$$

Recall that $\bar{i}$ and $\tilde{i}$ were defined in (9) and (10) respectively. With these notation, we can lower bound (130) as:

$$\sum_{k=1}^{K} \max\{0, \frac{1}{2}\log_2(P_k)\} - \max\{0, \frac{1}{2}\log_2\left(\frac{a_{\bar{i}} P_{\bar{i}}}{\max\{1, a_{\tilde{i}}\}}\right)\} - f_M(K) \tag{131}$$

As shown in [15], $M \leq 2K - 1$. Hence

$$f_M(K) \leq (2K-1)\left(\frac{K-1}{2} + \frac{K+1}{2}\log_2(K)\right) + \frac{1}{2}\log_2(K) \tag{132}$$

Applying it to (131), we get the theorem. ∎

## V. Upper Bound on the Secrecy Sum Rate

Let $n$ be the total number of channel uses. Define $V^n$ as: $V^n = \sum_{i=1}^{K-1} \sqrt{a_i} X_i^n + Z_K^n$. We use the shorthand $A_{a_1, a_2, \ldots, a_k}$ to denote $\{A_{a_1}, \ldots, A_{a_k}\}$. Then we have the following lemma:

*Lemma 6:*

$$nR^s_\Sigma \leq I\left(W_{1,\ldots,K-1}; Y^n_{1,\ldots,K-1}\right) - I\left(W_{1,\ldots,K-1}; V^n\right) + I\left(X^n_K; Y^n_K | X^n_{1,\ldots,K-1}\right) + n\varepsilon \tag{133}$$

where $\varepsilon$ is nonnegative and $\lim_{n \to \infty} \varepsilon = 0$.

*Remark 3:* Lemma 6 is an extension of the technique from [7]. The technique expresses the upper bound on the secrecy sum rate in two terms, as shown by (133). The second term in (133) corresponds to the point-to-point link between the $K$th user and its receiver. Interference is removed since $X_{1,\ldots,K-1}^n$ appears on the condition term. The first term in (133), as we will see later, can be bounded use the technique from [8], as it shares the same form as the secrecy sum rate upper bound for a multiple access wiretap channel considered therein, whose main channel are composed of the links between the first $K-1$ users and whose eavesdropper receives $V^n$, which is the interference experienced by the $K$th user. □

*Proof:* The two user case ($K=2$) has been shown in [7, Appendix]. The same technique can be used here to prove Lemma 6. We first prove

$$nR_\Sigma^s - n\varepsilon \leq I\left(W_{1,\ldots,K-1}; Y_{1,\ldots,K-1}^n\right) - I\left(W_{1,\ldots,K-1}; V^n\right)$$
$$+ I\left(W_{1,\ldots,K-1}; V^n|Y_K^n\right) + I\left(X_K^n; Y_K^n\right) \tag{134}$$

This can be done by starting from [7, (41)], with $W_1$ being replaced by $W_{1,\ldots,K-1}$, $Y_1$ being replaced by $Y_{1,\ldots,K-1}$, $X_1$ being replaced by $X_{1,\ldots,K-1}$, $Y_2$ being replaced by $Y_K$. The $V_1^n$ therein is replaced by $V^n$. In this way, the secrecy sum rate is upper bounded as:

$$R_\Sigma^s - n\varepsilon \tag{135}$$
$$= H(W_{1,\ldots,K}) - n\varepsilon \tag{136}$$
$$\leq I\left(W_{1,\ldots,K-1}; Y_{1,\ldots,K-1}^n\right) - I\left(W_{1,\ldots,K-1}; Y_K^n\right) + I\left(W_K; Y_K^n\right) \tag{137}$$
$$= I\left(W_{1,\ldots,K-1}; Y_{1,\ldots,K-1}^n\right) - I\left(W_{1,\ldots,K-1}; V^n\right) - I\left(W_{1,\ldots,K-1}; Y_K^n|V^n\right) \tag{138}$$
$$+ I\left(W_{1,\ldots,K-1}; V^n|Y_K^n\right) + I\left(W_K; Y_K^n\right) \tag{139}$$
$$\leq I\left(W_{1,\ldots,K-1}; Y_{1,\ldots,K-1}^n\right) - I\left(W_{1,\ldots,K-1}; V^n\right) + I\left(W_{1,\ldots,K-1}; V^n|Y_K^n\right) + I\left(W_K; Y_K^n\right) \tag{140}$$
$$\leq I\left(W_{1,\ldots,K-1}; Y_{1,\ldots,K-1}^n\right) - I\left(W_{1,\ldots,K-1}; V^n\right) + I\left(W_{1,\ldots,K-1}; V^n|Y_K^n\right) + I\left(X_K^n; Y_K^n\right) \tag{141}$$

Hence we have proved (134). It can be shown that the following inequality holds

$$I\left(W_{1,\ldots,K-1}; V^n|Y_K^n\right) + I\left(X_K^n; Y_K^n\right) \leq I\left(X_K^n; Y_K^n|X_{1,\ldots,K-1}^n\right) \tag{142}$$

by using the following steps:

$$I\left(W_{1,\ldots,K-1}; V^n|Y_K^n\right) + I\left(X_K^n; Y_K^n\right) \tag{143}$$

$$\leq I\left(X^n_{1,...,K-1}; V^n | Y^n_K\right) + I\left(X^n_K; Y^n_K\right) \tag{144}$$

$$= I\left(X^n_K; Y^n_K | X^n_{1,..,K-1}\right) + I\left(X^n_K; X^n_{1,...,K-1}\right)$$
$$- I\left(X^n_K; X^n_{1,...,K-1} | Y^n_K\right) + I\left(X^n_{1,...,K-1}; V^n | Y^n_K\right) \tag{145}$$

$$= I\left(X^n_K; Y^n_K | X^n_{1,..,K-1}\right) + I\left(X^n_{1,...,K-1}; V^n | Y^n_K\right) - I\left(X^n_K; X^n_{1,...,K-1} | Y^n_K\right) \tag{146}$$

$$= I\left(X^n_K; Y^n_K | X^n_{1,..,K-1}\right) + I\left(X^n_{1,...,K-1}; V^n | V^n + X^n_K\right) - I\left(X^n_K; X^n_{1,...,K-1} | V^n + X^n_K\right) \tag{147}$$

$$= I\left(X^n_K; Y^n_K | X^n_{1,..,K-1}\right) + I\left(X^n_{1,...,K-1}; X^n_K | V^n + X^n_K\right) - I\left(X^n_K; X^n_{1,...,K-1} | V^n + X^n_K\right) \tag{148}$$

$$= I\left(X^n_K; Y^n_K | X^n_{1,..,K-1}\right) \tag{149}$$

which yields (142). Applying (142) to (134) yields (133) in the lemma. ∎

Let $\tilde{V}^n = \sum_{i=1}^{K-1} \sqrt{\frac{a_i}{c}} X^n_i + \sqrt{\frac{1}{c}} Z^n_K + \sqrt{1-\frac{1}{c}} \tilde{Z}^n_K$, where $c = \max\{1, a_i, i = 1, ..., K-1\}$. $\tilde{Z}^n_K$ is a length-$n$ vector that has the same distribution as $Z^n_K$ but is independent from $Z^n_K$. Then we have the following lemma:

*Lemma 7:*

$$R^s_\Sigma \leq \lim_{n\to\infty} \frac{1}{n}\left(\sum_{i=1}^{K-1} I(X^n_i; Y^n_i) - I(X^n_{1,...,K-1}; \tilde{V}^n)\right)$$
$$+ \lim_{n\to\infty} \frac{1}{n} I(X^n_K; Y^n_K | X^n_{1,...,K-1}) \tag{150}$$

*Proof:* Because $\tilde{V}^n$ is a degraded version of $V^n$, from Lemma 6 and data processing inequality, we have

$$nR^s_\Sigma \leq I\left(W_{1,...,K-1}; Y^n_{1,...,K-1}\right) - I\left(W_{1,...,K-1}; \tilde{V}^n\right) + I\left(X^n_K; Y^n_K | X^n_{1,...,K-1}\right) + n\varepsilon \tag{151}$$

where $\varepsilon \geq 0$ and $\lim_{n\to\infty} \varepsilon = 0$. Next, we extend the derivation in [8, (58),(65)-(68)] to the first two terms, by replacing $Y^n$ with $Y^n_{1,...,K-1}$. The derivation in [8, (58),(65)-(68)] corresponds to the case of $K-1 = 2$ here. In particular, if we define the notation $i^c = \{j : 1 \leq j \leq K-1, j \neq i\}$. then we can write

$$\sum_{i=1}^{K-1} I(X^n_i; Y^n_{1,...,K-1} | X^n_{i^c}, W_i) - I(X^n_{1,...,K-1}; \tilde{V}^n | W_{1,...,K-1}) \geq 0 \tag{152}$$

which corresponds to [8, (58)]. To prove (152), we start with the fact that $\frac{a_i}{c} \leq 1, \forall i$. Then, using $\frac{a_i}{c} \leq 1, \forall i$, we have:

$$I\left(X^n_i; Y^n_{1,...,K-1} | X^n_{i^c}, W_i\right) \geq I\left(X^n_i; \tilde{V}^n | X^n_{i^c}, W_i\right) \tag{153}$$

The right hand side of (153) can be lower bounded as:

$$I\left(X_i^n; \tilde{V}^n | X_{i^c}^n, W_i\right) \tag{154}$$

$$=I\left(X_i^n; \tilde{V}^n | X_{1,...,i-1}^n, W_{1,...,K-1}, X_{i+1,..K-1}^n\right) \tag{155}$$

$$=I\left(X_i^n; \tilde{V}^n, X_{i+1,..K-1}^n | X_{1,...,i-1}^n, W_{1,...,K-1}\right) - I\left(X_i^n; X_{i+1,..K-1}^n | X_{1,...,i-1}^n, W_{1,...,K-1}\right) \tag{156}$$

$$=I\left(X_i^n; \tilde{V}^n, X_{i+1,..K-1}^n | X_{1,...,i-1}^n, W_{1,...,K-1}\right) \tag{157}$$

$$\geq I\left(X_i^n; \tilde{V}^n | X_{1,...,i-1}^n, W_{1,...,K-1}\right) \tag{158}$$

(157) follows because each user does its encoding independently. The message of each user is also independent from each other. Hence $I\left(X_i^n; X_{i+1,..K-1}^n | X_{1,...,i-1}^n, W_{1,...,K-1}\right) = 0$. From (153), (154)-(158), we have

$$I\left(X_i^n; Y_{1,...,K-1}^n | X_{i^c}^n, W_i\right) \geq I\left(X_i^n; \tilde{V}^n | X_{1,...,i-1}^n, W_{1,...,K-1}\right) \tag{159}$$

Adding (159) for $i = 1, ..., K-1$, we have (152).

We next use the fact that:

$$I(W_{1,...,K-1}; Y_{1,...,K-1}^n) - I(W_{1,...,K-1}; \tilde{V}^n)$$
$$\leq \sum_{i=1}^{K-1} I(W_i; Y_{1,...,K-1}^n | X_{i^c}^n) - I(W_{1,...,K-1}; \tilde{V}^n) \tag{160}$$

which corresponds to [8, (66)]. (160) can be proved as follows:

$$I\left(W_{1,...,K-1}; Y_{1,...,K-1}^n\right) - I\left(W_{1,...,K-1}; \tilde{V}^n\right) \tag{161}$$

$$= \sum_{i=1}^{K-1} I\left(W_i; Y_{1,...,K-1}^n | W_{1,...,i-1}\right) - I\left(W_{1,...,K-1}; \tilde{V}^n\right) \tag{162}$$

$$\leq \sum_{i=1}^{K-1} I\left(W_i; X_{i^c}^n, Y_{1,...,K-1}^n | W_{1,...,i-1}\right) - I\left(W_{1,...,K-1}; \tilde{V}^n\right) \tag{163}$$

$$= \sum_{i=1}^{K-1} \left(I\left(W_i; X_{i^c}^n | W_{1,...,i-1}\right) + I\left(W_i; Y_{1,...,K-1}^n | X_{i^c}^n, W_{1,...,i-1}\right)\right) - I\left(W_{1,...,K-1}; \tilde{V}^n\right) \tag{164}$$

$$= \sum_{i=1}^{K-1} I\left(W_i; Y_{1,...,K-1}^n | X_{i^c}^n\right) - I\left(W_{1,...,K-1}; \tilde{V}^n\right) \tag{165}$$

The lemma follows by adding the left hand side of (152) to the right hand side of (160) and using the fact

$$I\left(X_i^n; Y_{1,...,K-1}^n | X_{i^c}^n\right) = I\left(X_i^n; Y_i^n\right) \tag{166}$$

We next prove Theorem 2.

*Proof of Theorem 2:* The theorem follows by evaluating the bound in Lemma 7. This is done by extending [8, Theorem 4]. [8, Theorem 4] corresponds to the case with $K-1=3$.

We start with

$$\sum_{i=1}^{K-1} I(X_i^n; Y_i^n) - I\left(X_{1,...,K-1}^n; \tilde{V}^n\right) \tag{167}$$

$$= \sum_{i=1}^{K-1} h(X_i^n + Z_i^n) - h\left(\sum_{i=1}^{K-1} \sqrt{v_i} X_i^n + Z_K^n\right) - \sum_{i=1}^{K-1} h(Z_i^n) + h(Z_K^n) \tag{168}$$

We next derive an upper bound on the first two terms in (168). The main technique is the generalized entropy power inequality [8], [20]: Let $A_i^n, i=1,...,m$ be $m$ length-$n$ continuous random vector. Let $S_1,...,S_p$ be $p$ arbitrary subset of $\{1,...,m\}$. Then

$$2^{\frac{2}{n}h\left(\sum_{i=1}^{m} A_i^n\right)} \geq \frac{1}{\gamma}\left(\sum_{k=1}^{p} 2^{\frac{2}{n}h\left(\sum_{j \in S_k} A_j^n\right)}\right) \tag{169}$$

where $\gamma$ is the maximal number of sets in $S_1,...,S_p$ in which any $A_i^n$ appears.

Without loss of generality, we assume $v_1 \leq v_2 \leq ... \leq v_{K-1} \leq 1$. Define $v_K = 1, v_0 = 0$. Define $N_j^n, j=1,...,K$ as zero mean independent length-$n$ Gaussian vector whose component has unit variance. Then we have

$$2^{\frac{2}{n}h\left(\sum_{i=1}^{K-1} \sqrt{v_i} X_i^n + Z_K^n\right)} \tag{170}$$

$$= 2^{\frac{2}{n}h\left(\sum_{i=1}^{K-1} \sqrt{v_i} X_i^n + \sum_{j=1}^{K} \sqrt{v_j - v_{j-1}} N_j^n\right)} \tag{171}$$

$$\geq \frac{1}{K-1}\left(\sum_{i=1}^{K-1} 2^{\frac{2}{n}h\left(\sqrt{v_i} X_i^n + \sum_{j=1}^{i} \sqrt{v_j - v_{j-1}} N_j^n\right)} + \sum_{i=2}^{K} 2^{\frac{2}{n}h\left(\sum_{j=i}^{K} \sqrt{v_j - v_{j-1}} N_j^n\right)}\right) \tag{172}$$

$$= \frac{1}{K-1}\left(\sum_{i=1}^{K-1} 2^{\frac{2}{n}h\left(\sqrt{v_i} X_i^n + \sqrt{v_i} Z_i^n\right)} + \sum_{i=2}^{K} 2^{\frac{2}{n}h\left(\sum_{j=i}^{K} \sqrt{v_j - v_{j-1}} N_j^n\right)}\right) \tag{173}$$

$$= \frac{1}{K-1}\left(\sum_{i=1}^{K-1} v_i 2^{\frac{2}{n}h(X_i^n + Z_i^n)} + \sum_{i=2}^{K} 2\pi e(1 - v_{i-1})\right) \tag{174}$$

The general power of entropy inequality was used to obtain (172). In (173), we replace

$$\sum_{j=1}^{i} \sqrt{v_j - v_{j-1}} N_j^n \tag{175}$$

with $\sqrt{v_i}Z_i^n$, since these two terms have the same distribution.

Hence we have

$$\sum_{i=1}^{K-1} h\left(X_i^n + Z_i^n\right) - h\left(\sum_{i=1}^{K-1} \sqrt{v_i}X_i^n + Z_K^n\right) \tag{176}$$

$$\leq \sum_{i=1}^{K-1} h\left(X_i^n + Z_i^n\right) - \frac{n}{2}\log_2\left(\frac{1}{K-1}(\sum_{i=1}^{K-1} v_i 2^{\frac{2}{n}h(X_i^n + Z_i^n)} + \sum_{i=2}^{K} 2\pi e\left(1 - v_{i-1}\right))\right) \tag{177}$$

For any $i$, $1 \leq i \leq K-1$, (177) is a monotonically increasing function of $h(X_i^n + Z_i^n)$. Hence (177) is maximized when $X_i^n$ is chosen to have a Gaussian distribution with independent components. Each component is chosen to have the maximal possible variance $P_i$. Applying this result back to (168), we find it to be upper bounded by:

$$\sum_{i=1}^{K-1} C\left(P_i\right) - C\left(\frac{\sum_{i=1}^{K-1} v_i P_i}{K-1}\right) \tag{178}$$

This, along with the fact that $I\left(X_K^n; Y_K^n | X_{1,...,K-1}^n\right) \leq nC(P_K)$, gives us the result in the theorem.

■

## VI. CONCLUSION

In this work, we have considered the $K$-user ($K \geq 3$) Gaussian many-to-one interference channel with confidential messages. We derived the achievable secrecy sum rate as well as the upper bound on the secrecy sum rate. The achievable rate was obtained using layered coding and using nested lattice codes for each layer. The upper bound and the achievable secrecy sum rate matches in terms of secure degree of freedom. Since it also matches the degree of freedom of the sum rate when the secrecy constraints are removed, we observe the secrecy requirement does not reduce degree of freedom in this model.

We have also identified two cases where the gap between the upper bound and lower bound is only a function of $K$, and is independent of channel gains. One case is when the channel gains of the interfering links are all $\leq 1$ (direct link gains). The other case is when the power constraints of the first $K-1$ users at the $K$th receiver are the same and the channel gains of the interfering links are the same.

Nested lattice codes we used in this work has recently found many applications in solving information theoretic secrecy problems, see [21] and [21] for examples where the equivocation are bounded in terms of Shannon entropy and Rényi entropy respectively. In [21], we use nested lattice codes to prove that Gaussian signaling is suboptimal at high SNR for a large class of two-user Gaussian channel with secrecy constraints. This work can be viewed as a generalization of [21] to the $K$-user case ($K \geq 3$).